\documentclass[
aip,
pop,
amsmath,amssymb,
preprint,
]{revtex4-2} 

\preprint{AIP/123-QED}

\usepackage[T1]{fontenc}
\usepackage[english]{babel}

\usepackage[group-minimum-digits=4,group-separator={,},group-digits=integer]{siunitx}
\DeclareSIUnit{\angstrom}{\textup{\AA}} 

\usepackage{graphicx}
\usepackage{xcolor}
\graphicspath{{./Figures/}}  
\usepackage[caption=false]{subfig}
\usepackage{dcolumn}
\usepackage{bm}

\usepackage{booktabs,makecell,multirow}

\usepackage{nicefrac}

\usepackage[version=4]{mhchem}

\newcommand{\bv}{\bm{v}}
\newcommand{\br}{\bm{r}}
\renewcommand\epsilon{\varepsilon}

\newcommand{\refEq}[1]{Eq.~(\ref{#1})}
\newcommand{\refTab}[1]{Table~\ref{#1}}
\newcommand{\refFig}[1]{Fig.~\ref{#1}}

\newcommand{\cfS}[1]{Sect.~\ref{#1}}

\begin{document}

\title{Self-Sputtering of the Lennard-Jones Crystal}

\author{Nicolas A. Mauchamp}
\email{mauchamp@ppl.eng.osaka-u.ac.jp}
\author{Kazumasa Ikuse}
\author{Michiro Isobe}
\author{Satoshi Hamaguchi}
\email{hamaguch@ppl.eng.osaka-u.ac.jp}
\affiliation{Center for Atomic and Molecular Technologies, Graduate School of Engineering, Osaka University, 2-1 Yamadaoka, Suita, Osaka 565-0871, Japan}

\date{\today}

\begin{abstract}
The self-sputtering yield of the (100) face-centered cubic (fcc) crystal surface consisting of particles interacting with the Lennard-Jones (LJ) potential is presented as a function of the normalized incident particle kinetic energy for normal incidence. Because the self-sputtering yield depends only on the normalized incident energy, the yield curve presented here is the universal curve, independent of the Lennard-Jones parameters, and therefore serves as the fundamental reference data for the LJ system. The self-sputtering yield data are also compared with experimentally obtained self-sputtering yields of some metals, which shows reasonable agreement at relatively low ion incident energy where mostly deposition occurs. At higher ion energy, the self-sputtering of such an LJ material does not represent those of real solids. This is because the repulsive interactions of the LJ potential do not represent those of actual atoms at short distances. The angle dependence of the self-sputtering yield is also presented for some selected normalized energies.

\medskip
The following article has been submitted to Physics of Plasmas. After it is published, it will be found at https://aip.scitation.org/journal/php.
\end{abstract}

\maketitle 

\section{Introduction} 
\label{S::introduction}

Sputtering\cite{citeBook_Sigmund06,citeBook_Eckstein07} is a process of removing atoms from a solid-material surface by the impact of energetic incident ions or atoms. Such an impact causes a collision cascade of atoms inside the material, where billiard-like multiple collisions can eject some atoms from its surface. When the incident ions or atoms are of the same species as those of the atoms constituting the surface material, the sputtering is called the self-sputtering. 

Sputtering was first observed in 1853 by Grove\cite {grove:1853aa} in a cathode tube experiment.
In 1891, about 40 years later, Crooks\cite{crookes:1891aa} studied some properties of sputtering, which he called ``electrical evaporation'', for several different materials.
The first realistic mathematical model of sputtering was proposed by Sigmund\cite{sigmund_PR69_sputter_theory} in 1969, nearly 80 years after the study by Crooks.  
After the study by Sigmund, many studies, both theoretical and experimental, followed and contributed to a better understanding of the sputtering mechanisms. However, sputtering phenomena are so complex that no mathematical formula to predict an experimentally observable quantity such as the sputtering yield has been derived from the first principles. 
When first observed in 1853, sputtering was considered as an unwanted effect causing damages on the electrodes and contamination of the plasma.
Over time, many applications were found and sputtering became a major tool for surface modification, such as surface etching, thin film deposition, and surface cleaning. Especially since the advent of mass production of large-scale-integration (LSI) devices, plasmas have been used for the processing of semiconductor devices.\cite{citeBook_Chapman80,roadmap17,Weltmann19}  
Among various plasma processing, etching processes \cite{Oehrlein_18PSST,Hamaguchi_IBM} often employ sputtering techniques combined with surface chemistry, which are called chemical sputtering or reactive ion etching. Plasma processing technologies have contributed significantly to the miniaturization of semiconductor devices as it is the main driving technology for fabricating  micro-/nano-scale structures on a semiconductor material surface. \cite{Cotler_JVSTB87_profile_sim,McVittie_profile_sim_JAP89,dalvie92_IEEE,92JVSTB_Singh_Saqfeh_McVittie_profile_sim,Hamaguchi_math_shape_96,Hoekstra_Kushner_JVSTA97_surface_profile_sim,Mayo97,Kushner_profile_sim_JVSTA20} 

One of the main quantities characterizing sputtering phenomena is the sputtering yield, i.e., the average number of atoms removed from the surface at a single impact of an ion or an atom. For self-sputtering, if the sputtering yield is larger than unity, the surface is etched; if it is less than unity, the material is being deposited.
The sputtering yield in general depends on the surface conditions.
When a clean surface is exposed to a steady ion beam, the initial sputtering yield can be different from those in later time, when the surface becomes rough and/or the surface chemical compositions change due to the deposition of incident species.
In this article, unless otherwise stated, we refer to the sputtering yield as that in steady state, where the surface is exposed to the steady incident ions for a sufficiently long time and its surface roughness and chemical compositions no longer change in time.

As one expects, the sputtering yield depends on the incident species and their kinetic energy. Experimentally, the sputtering yield can be measured in ion beam experiments.\cite{Karahashi14_review} Earlier sputtering yield data for single-element materials by single-element ions were extensively compiled in Refs.~\citenum{citeBook_Eckstein07} and \citenum{Yamamura_Tawara_ADNDT96}.
More recently, sputtering yields were also measured with mass and energy controlled ion beams experiments for \ce{Si}- and \ce{C}-based materials, \cite{Tachi_1981,Zalm83,Tachi_1986, Ishikawa_03, Karahashi_04,Ito11a,Ito13JVST,Yanai_05,Yoshimura_PMMA_12,Yoshimura_PMMA_photon_hydronge_13,Karahashi14_review,Karahashi_2017} metals,\cite{Ikuse_Au_09,Satake_Ta_15,Li_Ta_15} and metal oxides.\cite{hine_MgO_07,hine_MgO_inert_08,Yoshimura_CaO_SrO_BaO_12,Ikuse__Mg(OH)2_12,Li_ITO_15,Li_ZnO_16,Li_ZnO_Heffects_17}
Sputtering can cause damages on the  material surface, which have also been extensively studied.\cite{Yabumoto_JJAP81_surface_damage,Pang_plasma_damage_JAP83,Ohchi_Si_damage_08, Ito11b,Ito_2012,Hirata_JJAP17_ITO_damage,Li_ITO_Hdope_18}
   
Sputtering yields can be evaluated theoretically. For example, at relatively high ion incident energy, numerical simulations based on binary-collision models\cite{Biersack1980,Ziegler1985,Ziegler1991} can be used to evaluate the sputtering yields of various materials.
Under more general conditions, classical molecular dynamics (MD) simulation, where the motions of atoms interacting among themselves are analyzed dynamically with the predetermined interatomic potential models, can be used to estimate the sputtering yields.\cite{Graves_review_JPD09}
Unlike binary-collision models, MD simulation can self-consistently model multi-body collisions, which play the dominant role when atoms with relatively low kinetic energies collide with other atoms in a material.
MD simulations have been widely used to study sputtering and deposition of \ce{Si}-based materials\cite{Stansfield1989,Smith1989,schoolcraft1991,Carter92,Feil93,barone95,Kubota_Economou1998,Hamaguchi_Si_SiO2_02,Yamada04,Taguchi07,Tinacba19,Tinacba_SF5_21} and metals.\cite{WEBB1984,Hansen:1999,Hanson:2001,Stoller:2016aa,METSPALU_2018,Mauchamp_Ni_EAM_21} 

Although physical phenomena of sputtering are highly complex and the exact sputtering yields may depend sensitively on the nature of atomic interactions of the system, it is also known empirically that the sputtering yield values can be approximately estimated with a relatively small number of physical characteristics of the system, e.g., the incident ion energy, incident ion mass, the atomic mass of the surface material, surface binding energy of the material, etc.\cite{Kino_PoP2021}
This suggests that relatively simple models of atomic interactions account for a large part of the sputtering phenomena involving those atoms.

One of the simplest interatomic potential models is the 12-6 Lennard-Jones (LJ) potential.
It has been widely used to study material properties of face-centered cubic (fcc) crystals, transport properties and equation of state\cite{Johnson_LJ_EOS_93} for fluids and gases, and the phase transitions among them.
The first numerical simulation to examine thermodynamical properties of the system of LJ particles (i.e., particles interacting with the LJ potential) was performed by Wood and Parker.\cite{wood_LJ_MC_57}
The phase diagram of the LJ system (i.e., the system consisting of LJ particles) is given in Ref.\citenum{Bringa_LJ_MD_phasediagram_sputter1_98}.     

Although MD simulation has been used to study thermodynamical properties of the system consisting of LJ particles, 
there seems few studies that determines the sputtering yield of an LJ solid (i.e., a solid consisting of LJ particles) as a function of the incident particle energy.
For example, Anders {\it et al.} studied self sputtering by cluster particles, including single particles, for a clean surface of an fcc crystal consisting of modified LJ particles.\cite{PRB_anders_cluster_04} Electronic sputtering (i.e., sputtering by energetic electrons) have been also studied for LJ systems. \cite{Bringa_LJ_MD_phasediagram_sputter1_98,Bringa_LJ_MD_yield_sputter2_99,Bringa_electronic_sputter_PRB_99}
However, it seems no systematic study was performed to determine the steady-state self-sputtering yield of a genuine LJ solid subject to an energetic single particle beam prior to our study presented here.
In our study, we use MD simulations to determine the self-sputtering yield of an fcc LJ crystal as a function of the incident particle kinetic energy and examine how it defers from the experimentally observed sputtering yields of various materials. 

The rest of this article is structured as follows: \cfS{S::LJpot} presents the governing equations solved by MD simulation.  
\cfS{S::MDsim} gives the simulation method.
In \cfS{S::self-sputter results}, the simulation results, including those by modified LJ potentials, are presented and the comparison with the self-sputtering yields of some metals are discussed. 
Finally, in \cfS{S::Concl}, the conclusion is presented.

\section{\label{S::LJpot}L-J potential function and equations of motion}

The LJ potential function $V_\text{LJ}$ between two particles is commonly expressed as

\begin{equation}
\label{Eq::LJpotential}
    V_\text{LJ} (r) = 4 \epsilon \left[ \left(\frac{\sigma}{r}\right) ^ {12} - \left( \frac{\sigma}{r}\right) ^ 6 \right],
\end{equation}

where $r$ is the inter-particle distance. 
The LJ potential depends on two parameters:
the dispersion energy $\epsilon$, which characterizes the binding energy between the two interacting particles, and
the particle radius $\sigma$, which corresponds to the inter-particle distance where the potential is null.

Using the following normalized variables

\begin{equation}
\xi = \frac{r}{\sigma}
\quad  \text{and} \quad
\tilde{V} = \frac{V_\text{LJ}}{4\epsilon},
\label{E_norm1}
\end{equation}

the normalized LJ potential function can be written as

\begin{equation}
\label{E_LJ_eqn_Norm}
\tilde{V} (\xi) = \left[ \frac{1}{\xi ^ {12}} - \frac{1}{\xi ^ 6} \right] \quad.
\end{equation}

\begin{figure}[htbp]
\begin{center}
\includegraphics[width=\linewidth]{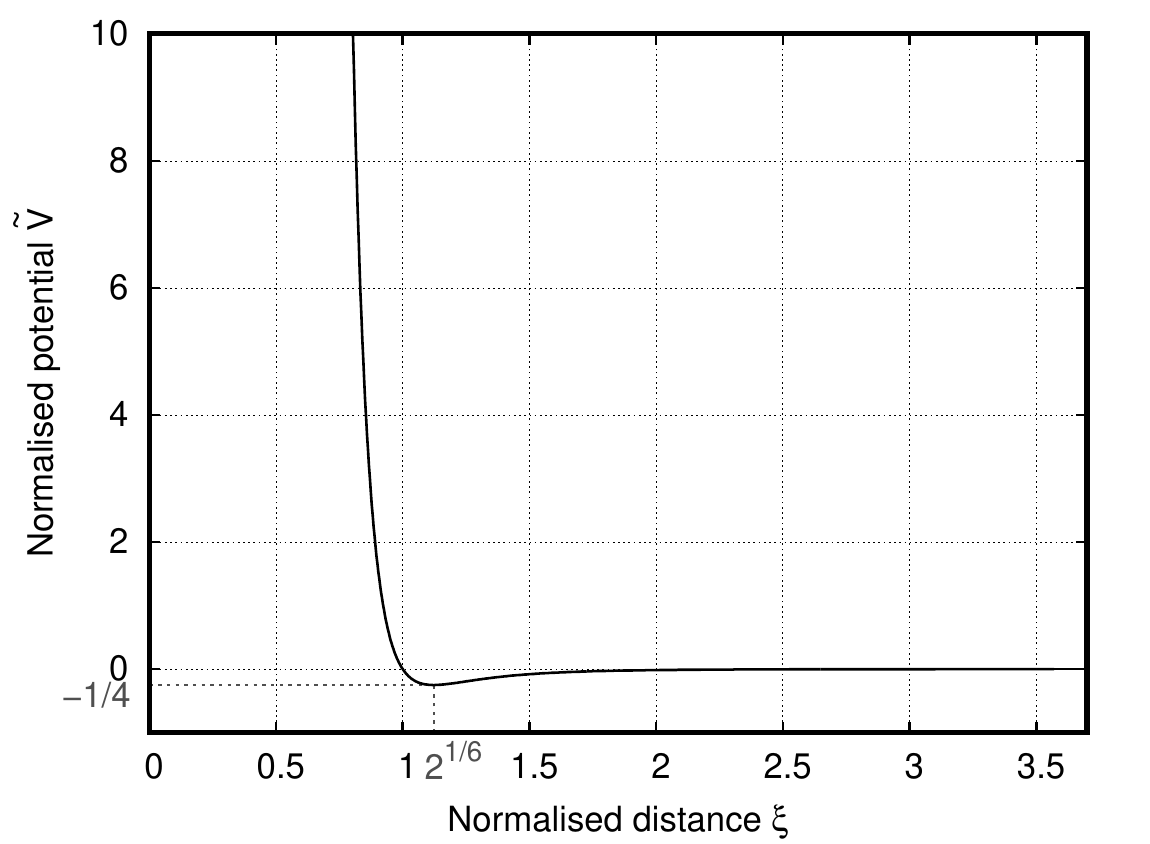}
\caption{Normalized 12-6 LJ potential function $\tilde{V}$ as a function of the normalized distance $\xi$.
The potential takes its minimum $\tilde{V}_\text{min} = -1/4$ at $\xi_\text{min} = 2 ^ {1/6}$ and is null for $\xi_0 = 1$.
In the MD simulations used in this study, the potential function is cut at $\xi_\text{C} \simeq \num{3.7}$.}
\label{F::LJpot_Norm1}
\end{center}
\end{figure}

The normalized potential function presents two characteristic points:
the potential $\tilde{V}$ is null for $\xi_0 = 1$ 
and
takes its minimum value $\tilde{V}_\text{min} = -1/4$ at $\xi_\text{min} = 2 ^ {1 / 6}$.
In the MD simulations performed for this study, the potential function was cut at $\xi_\text{C} \simeq \num{3.7}$.
The value of normalized potential $\tilde{V}$ at this cutoff distance is so small  that it has essentially no effect on the dynamics of LJ particles. 
The normalized potential function $\tilde{V}$ is shown in \refFig{F::LJpot_Norm1}. 

The equations of motion for the LJ system consisting of $N$ particles are given by

\begin{equation}
\label{E::EqMotion}
m \frac{d \bv_i}{d t} =
- 4 \epsilon
\frac{\partial}{\partial \br_i} 
\sum_{j (\neq i)} 
\left[ \left( \frac{\sigma}{r_{ij}} \right) ^ {12} - \left( \frac{\sigma}{r_{ij}} \right) ^ {6} \right]
\end{equation}

Here $\br_i$ and $\bv_i$ denote the position and velocity of the $i$-th particle ($i = 1, 2, \dots N$) and the distance between the $i$-th and $j$-th particles is represented by $r_{ij} = \lvert \br_i - \br_j \rvert$.
$\sum_{j (\neq i)}$ denotes the summation over all particles ($j = 1,2, \cdots, N$) except for the $i$-th particle.
The mass of the LJ particle is denoted by $m$. To use the LJ system to examine sputtering phenomena, we assume that the LJ particle represents a single atom or ion. 
In this study, no Coulomb interaction is considered. In actual experiments, incident ions are considered to be charge-neutralized by an Auger emission process right before their impact on the surface. Based on this premise,
we assume that both atoms and ions are represented by the same LJ particle and governed by Eq.~(\ref{E::EqMotion}) for the sake of simplicity.

With the normalization similar to \refEq{E_norm1}, i.e.,  

\begin{equation*}
    \bm{\xi}_i = \frac{\br_i}{\sigma}, \quad
    \bar{\bm{v}_i} = \frac{d \bm{\xi}_i}{d\tau} = \bv_i \frac{T_0}{\sigma} \quad \text{and} \quad  
    \tau = \frac{t}{T_0}  \quad  \text {with} \quad 
    T_0  = \frac{\sigma}{2}\sqrt{\frac{m}{\epsilon}}, 
\end{equation*}

\refEq{E::EqMotion} may be normalized as

\begin{equation}
\label{E::EqMotion_Norm}
\frac{d \bar{\bm{v}}_i}{d \tau} =
-  \frac{\partial}{\partial \bm{\xi}_i} \sum_j\ \tilde{V}\left( \xi_{ij}\right), 
\end{equation}

where $\xi_{ij} = \lvert \bm{\xi}_i - \bm{\xi}_j \rvert$.
It should be noted that \refEq{E::EqMotion_Norm} has no free parameter. Therefore the dynamics of this system is determined only by the initial and boundary conditions.
For example, the self-sputtering yield of an LJ solid, which is a dimensionless quantity, depends only on the normalized incident energy and angle of incidence.
The normalized kinetic energy $\mathcal E$ of a particle is defined by $\mathcal E = \bar{v}^2 / 2$ with the normalized velocity $\bar{v} = vT_0/\sigma$. Here $v = \lvert \bv \rvert$ and the relation between $\mathcal E$ and the dimensional kinetic energy $E = mv^2/2$ is given by 

\begin{equation}
\label{E_nomarlized_energy}
\mathcal{E}= \frac{E}{4\epsilon} \,\, .
\end{equation}

\section{\label{S::MDsim}MD simulation} 
In this study, classical molecular dynamics (MD) simulations were used to investigate sputtering phenomena of the LJ system. 
All MD simulations in this work were performed with LAMMPS,\cite{plimpton:1995aa} a classical molecular dynamics simulation code distributed by Sandia National Laboratories, USA, under the GNU General Public License (GPL).

The material used in this study was made of a collection of $\num{27000}$ or $\num{122500}$ LJ particles (i.e., atoms) arranged in a fcc crystalline structure of lattice constant $l_a \simeq 1.54 \sigma$. 
It had a rectangular shape whose size in the $x$, $y$ and $z$ directions was
\num{15}, \num{15} and \num{30} times the lattice constant $l_a$ (\num{27000} atoms) if the normalized kinetic energy $\mathcal E$ of the incident particle satisfied $\mathcal E < \num{1.25e3}$.
If the incident energy was higher, i.e., $\mathcal E > \num{1.25e3}$, we employed a larger material of \num{25}, \num{25} and \num{50} times the lattice constant $l_a$ (\num{122500} atoms) 
to ensure that the material subject to the ion bombardment was large enough to incorporate sputtering phenomena with the corresponding large sputtering yields.
The material sizes used in this study were selected such that the sputtering yield obtained from MD simulations did not depend on the employed material size.
In all cases, the Miller index of the top surface was (100). 
This crystalline solid is known to be the most stable crystal for LJ particles at low temperature.\cite{Bringa_LJ_MD_phasediagram_sputter1_98}

This material was placed at the bottom of a simulation box that had the same horizontal dimensions (i.e., in the $x$ and $y$ directions) and a total height larger than the height of the material by $5 l_a$ to keep some space above the initial top surface for particle injections and surface dynamics.
In the horizontal directions, periodic boundary conditions were applied, such that the model surface could represent an infinitely large surface. 
The atoms in the few atomic layers at the bottom of the substrate were fixed in position to prevent the substrate from moving downward by ion impact on the top surface. Atoms leaving the top or bottom surface of the simulation box were considered to be lost and removed from the system. Since the substrate was deep enough, all atoms that left the system did so through the top surface of the simulation box, which means they were either sputtered atoms from the surface or reflected incident atoms or ions.  

We also performed MD simulations using larger and deeper materials to confirm that the sputtering yields obtained from MD simulations do not essentially depend on the material size. The material was initially thermalized at a normalized temperature of 
$\bar{T} = k_BT/4\epsilon = \num{6.5e-3}$, with $k_B$ being Boltzmann constant and $T$ being the temperature of the material. This initial temperature corresponds to $T = \SI{300}{\kelvin}$ for a typical material with $\epsilon = \SI{1}{\electronvolt}$.

\begin{figure}[htbp]
\begin{center}
\includegraphics[width=0.3\linewidth]{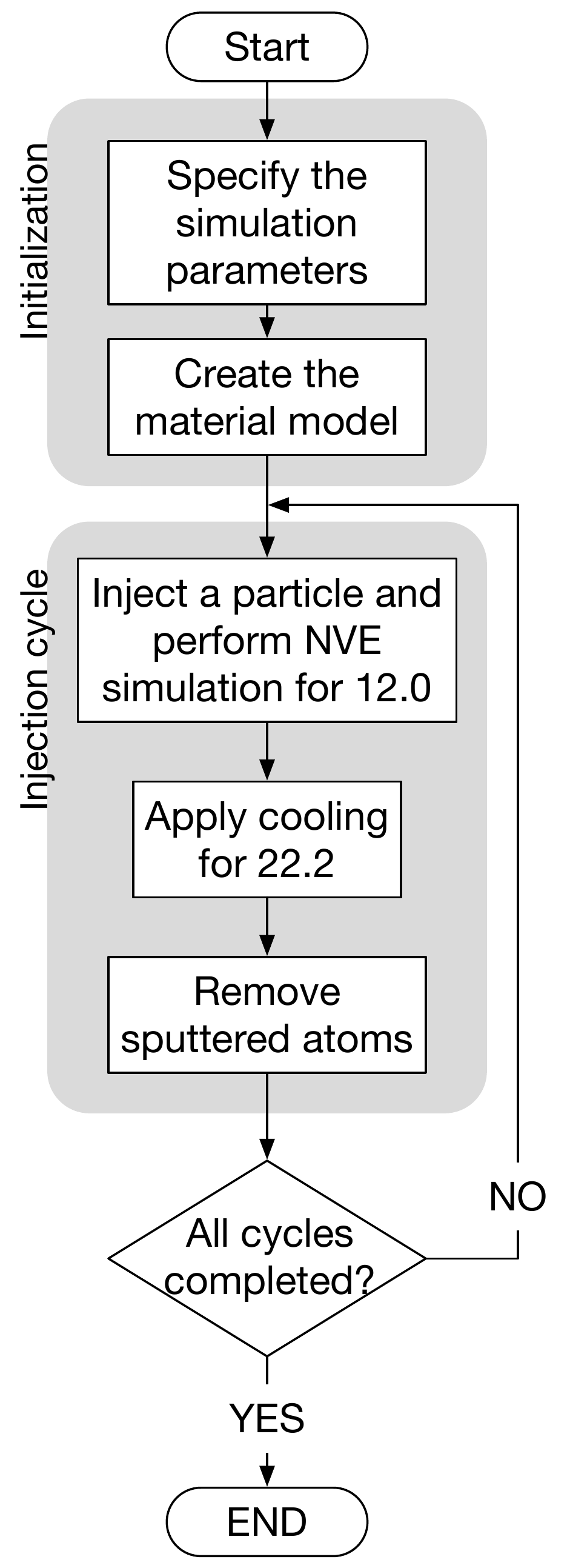}
\caption{Flow chart of the MD simulation.
Times are given in units of the normalized time $\tau$.}
\label{F::Flowchart}
\end{center}
\end{figure}

Sputtering simulations were performed in a manner similar to those of Ref.~\citenum{Mauchamp_Ni_EAM_21}. The flow chart of the simulation cycles is given in \refFig{F::Flowchart}. After setting up a material surface model 
at $\tau =0$, a single ion (or atom) was injected into the material surface at a random location of impact on the material surface with a given normalized incident energy $\mathcal{E}$. The incident angle $\alpha$ is defined with respect to the surface normal (i.e., the $z$ axis). Unless otherwise stated, the normal incidence angle, i.e., $\alpha = 0$, was selected. In our model, atoms and ions were treated equally and no Coulomb interaction was considered, as discussed earlier. 
The MD simulation was performed under the constant total-energy (NVE) conditions up to $\tau = 12.0$. 
The normalized time step $\delta\tau$ used to integrate the equations of motion was $\delta\tau = \num{1.70e-3}$ if the normalized incident particle energy $\mathcal E < \num{1.25e3}$, and $\delta\tau = \num{0.85e-3}$ if $\mathcal E > \num{1.25e3}$.

A Langevin cooling\cite{PRB_schneider:1978aa} was then applied from $\tau = 12.0$ to $30.8$ (for the duration of \SI{1.1}{\pico\second} under the typical material mentioned above), followed by the application of constant temperature (NVT) conditions from $\tau = 30.8$ to $34.2$ to emulate a slow-cooling and thermalization process to force the system to reach a thermalized state at $\bar{T} = \num{6.5e-3}$. Such a thermalization process would take place on much longer timescales in real material sputtering. 
The target temperature $\bar{T}_\text{target}$ and damping parameter $\tau_\text{damp}$ (i.e., inverse of the normalized friction coefficient) of the Langevin thermostat used in this study were $\bar{T}_\text{target} = \num{6.5e-3}$ and $\tau_\text{damp} =17.1$.
At $\tau = 34.2$, all desorbed atoms from the surface (i.e., atoms not bonded with the substrate) were considered as sputtered atoms, removed from the system, and a cycle of single ion injection was completed.
A summary of the simulation parameters with the corresponding dimensional values is given in \refTab{T::SimuPara}.

\begin{table}[htbp]
\caption{\label{T::SimuPara}Summary of the simulation parameters used in this study.
Physical quantities are given for $m = \SI{58.8}{\gram\per\mol}$ (i.e., the atomic mass of \ce{Ni}), $\epsilon = \SI{1}{\electronvolt}$, and $\sigma = \SI{1.5}{\angstrom}$. Here $\mathcal E$ and $E$ denote the normalized and dimensional kinetic energy values of the incident particle. 
}
\begin{ruledtabular}
\begin{tabular}{@{} l l c c @{}}
\multicolumn{2}{c}{Parameter} & Normalized & Physical\\
\midrule
\multirow{2}{*}{Timestep $\delta\tau$} & if $\mathcal E < \num{1.25e3}$ \,\,\,\, ($E < \SI{5000}{\electronvolt}$)  & \num{1.70e-3} & \SI{0.1}{\femto\second} \\
 & if $\mathcal E > \num{1.25e3}$ \,\,\,\, ($E > \SI{5000}{\electronvolt}$) &  \num{0.85e-3} & \SI{0.05}{\femto\second} \\
\multirow{2}{*}{Ion dose} & if $\mathcal E < \num{1.25e3}$ \,\,\,\, ($E < \SI{5000}{\electronvolt}$) & \num{1.87} & \SI{8.31e15}{\per\square\centi\meter} \\
 & if $\mathcal E > \num{1.25e3}$ \,\,\,\, ($E > \SI{5000}{\electronvolt}$) &  \num{0.67} & \SI{2.98e15}{\per\square\centi\meter} \\
 
 \multicolumn{2}{l}{Target temperature $\bar{T}_\text{target}$} & \num{6.5e-3} & \SI{300}{\kelvin} \\
\multicolumn{2}{l}{Damping parameter of the Langevin thermostat $\tau_\text{damp}$} & \num{17.1} & \SI{1.1}{\pico\second} \\
\multicolumn{2}{l}{Time for NVE simulation} & \numrange{0.0}{12.0} & \SIrange{0.0}{0.7}{\pico\second} \\
\multicolumn{2}{l}{Time for Langevin thermostat } & \numrange{12.0}{30.8} & \SIrange{0.7}{1.8}{\pico\second} \\
\multicolumn{2}{l}{Time for NVT simulation} & \numrange{30.8}{34.2} & \SIrange{1.8}{2.0}{\pico\second} \\

\end{tabular}
\end{ruledtabular}
\end{table}

This cycle was repeated many times until the average number of sputtered atoms per ion injection reached steady state, i.e., became independent of the ion dose.
In this study, the cycle was typically repeated \num{1 000} times (i.e., \num{1000} ion injections, which corresponds to a normalized ion dose of \num{1.87} in the case of $\num{270000}$ atoms or \num{0.67} in the case of $\num{122,500}$ atoms).
Typically the surface roughness due to the ion bombardment also reaches steady state when the sputtering yield no longer depends on the ion dose.

\section{\label{S::self-sputter results}Self-sputtering simulations} 
As discussed earlier, the sputtering yield $Y$ is expressed by

\begin{equation*}
Y = \frac{N_\text{Sputtered}}{N_\text{Incident}} ,
\end{equation*}

where $N_\text{Sputtered}$ and $N_\text{Incident}$ represent the total number of atoms leaving the surface and the total number of injected atoms over a specific time period.
If we select a sufficiently short time period to evaluate the sputtering yield, it is called an instantaneous sputtering yield.
As discussed earlier, because the surface roughness can change during the sputtering process, the instantaneous sputtering yield can evolve in time.
Typically the surface conditions reach steady state and the instantaneous sputtering yield fluctuates around a constant value after a sufficient ion dose.
In what follows, we refer to this constant value in steady state as the sputtering yield. 

In the case of self-sputtering, all sputtered atoms and incident ions or atoms are of the same kind and therefore indistinguishable macroscopically.
Therefore, all atoms leaving the material surface are counted in $N_\text{Sputtered}$, regardless of their origin, whether from the material or incident particles. 
(Computationally we could distinguish them, but we do not discuss the difference in this study.)
In this case, the threshold between deposition and sputtering is for $Y = 1$, where the number of atoms in the material does not change. 
For $Y < 1$, incident particles are deposited on the material surface and, for $Y > 1$, the material is being etched.

\begin{figure}[htbp]
\hfill
\subfloat[\label{F::Sim_boxA}]{%
\includegraphics[width=0.3\columnwidth]{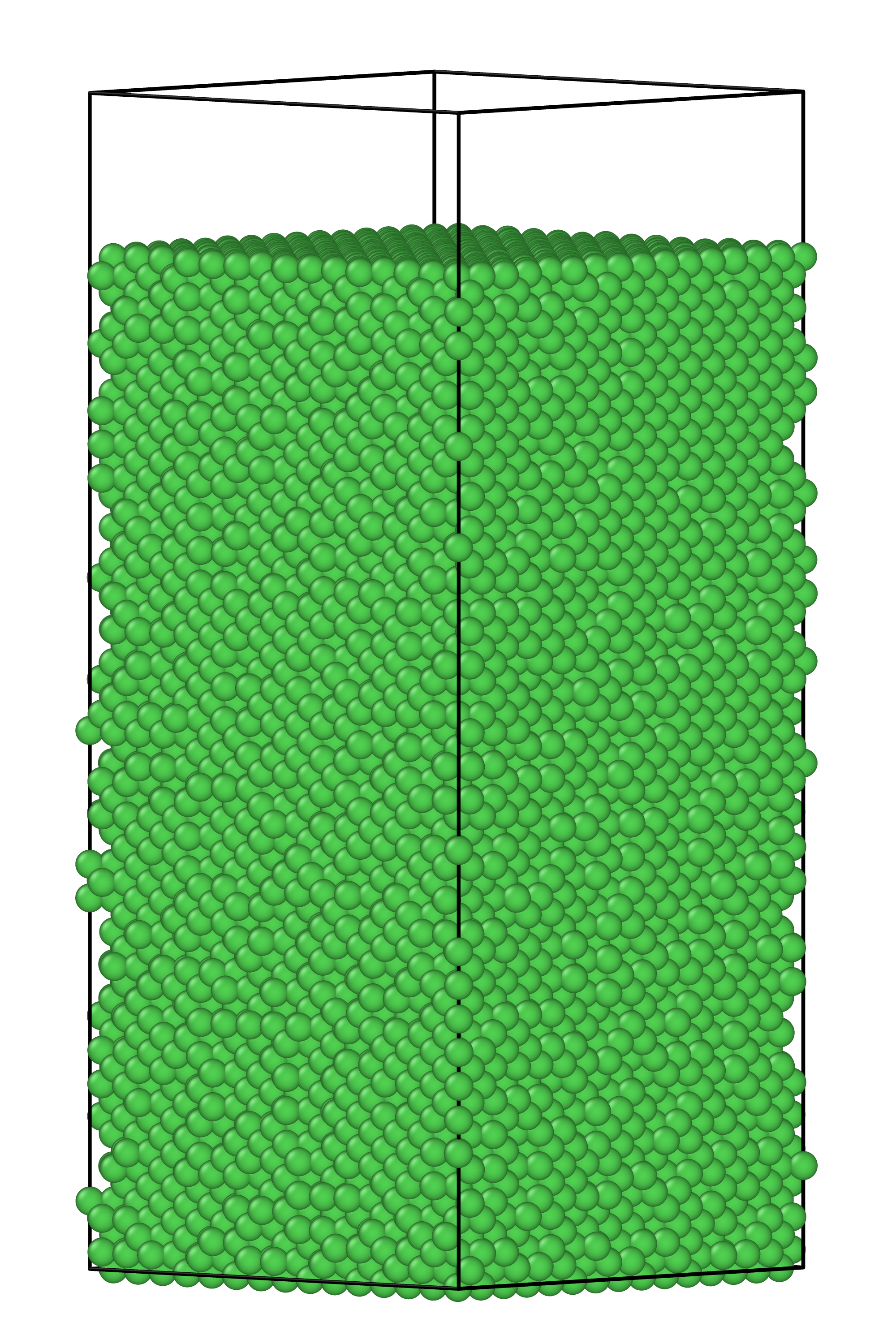}%
}\hfill
\subfloat[\label{F::Sim_boxB}]{%
\includegraphics[width=0.3\columnwidth]{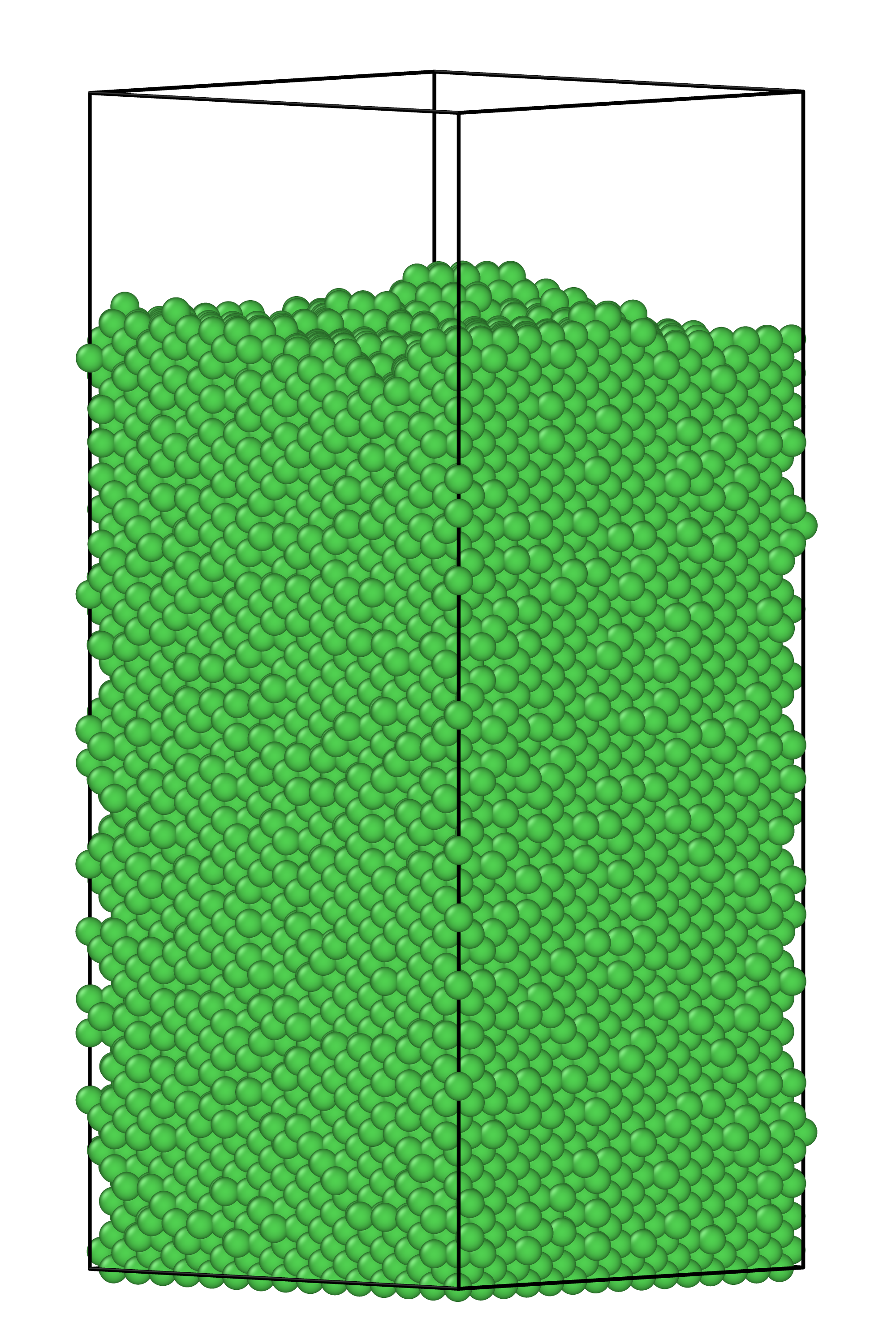}%
}\hfill\hfill
\caption{
Three-dimensional representations of the LJ material model used in this study.
Each green sphere represents an LJ particle and the black frame represents the simulation box. The periodic boundary conditions are imposed in the horizontal directions, so that the top surface represents an infinitely wide surface.
\protect\subref{F::Sim_boxA} The initial material model with the (100) top surface, which consists of \num{27000} particles. \protect\subref{F::Sim_boxB} The material model after \num{1000} LJ particle injections (i.e., a normalized ion dose of \num{1.87}) with a normalized incident energy of $\mathcal{E} = 250$.
Both material models are in thermal equilibrium at a normalized temperature $\bar{T} = \num{6.5e-3}$.
}
\label{F::Sim_box}
\end{figure}

The initial material model used in this study is shown in \refFig{F::Sim_boxA} and the material model after 1,000 particle injections with a normalized incident particle energy $\mathcal{E}$  of \num{250} is given in \refFig{F::Sim_boxB}.
It is seen that the surface receded and was roughened due to the sputtering.

\subsection{Energy and angle dependence of the self-sputtering yield} 

\begin{figure}[htbp]
\begin{center}
\includegraphics[width=\linewidth]{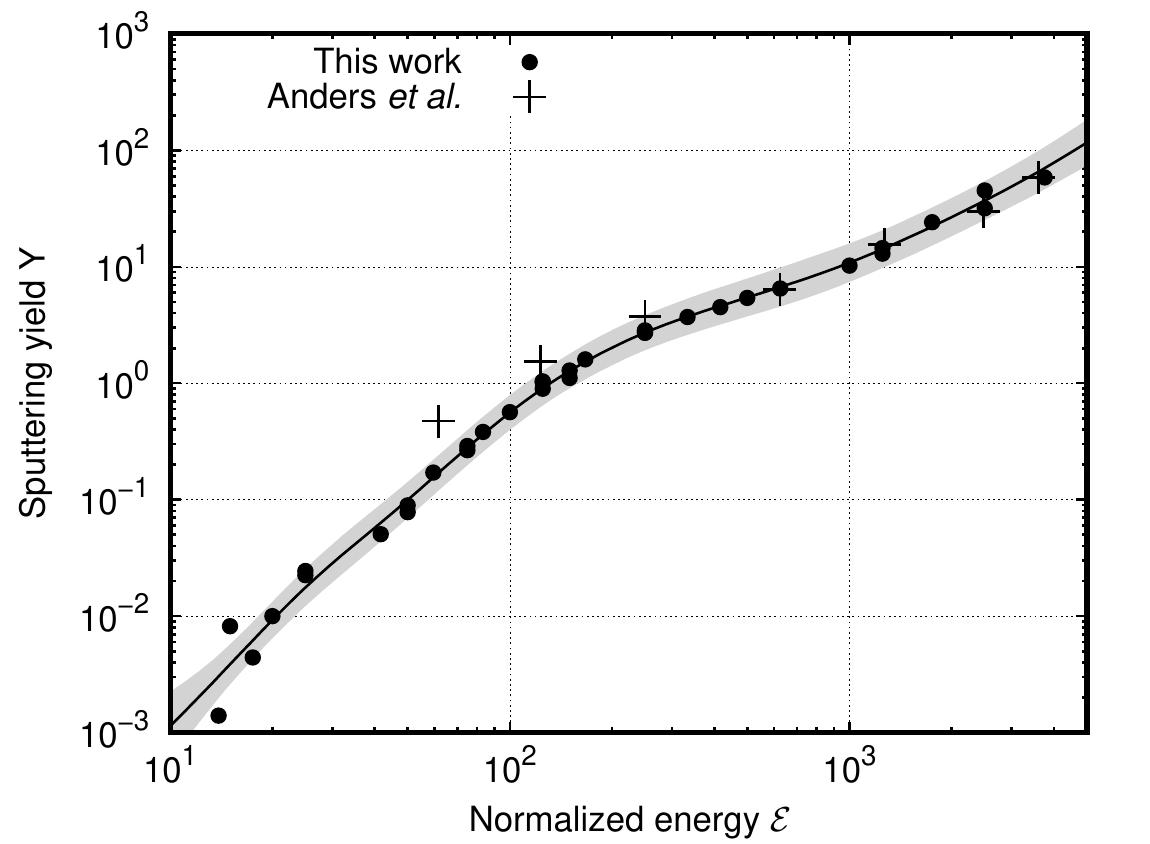}
\caption{Energetic dependency of the self-sputtering yield at normal incidence.
The filled circles represent the sputtering yields obtained from MD simulations of this study and the plus ($+$) signs are those obtained by Anders {\it et al.} of Ref.~\citenum{PRB_anders_cluster_04}. 
The solid curve is the prediction obtained using Gaussian Process Regression (GPR) and the gray area corresponds to the uncertainty of the prediction.
The predicted sputtering yield curve gives a threshold at $\mathcal{E} \simeq 132$, at which $Y = 1$. 
}
\label{F::EnergDepend}
\end{center}
\end{figure}

The dependency of the self-sputtering yield of the LJ system on the normalized incident kinetic energy obtained from MD simulation is given in \refFig{F::EnergDepend}.
The angle of incidence was normal to the material surface.
In this figure, the filled circles represent the sputtering yields obtained from MD simulations of this study and the plus ($+$) signs are those obtained by Anders {\it et al.} of Ref.~\citenum{PRB_anders_cluster_04}. 
The main difference between this study and Ref.~\citenum{PRB_anders_cluster_04} is that Anders {\it et al.} used a modified LJ potential, connected to the KrC potential\cite{wilson:1977aa} at short separation whereas this study shown in \refFig{F::EnergDepend} is based on the original LJ potential. Furthermore, Anders \textit{et al.} used a clean surface for every particle injection while, in the current work, the material surface is subject to successive particle injections and the self-sputtering yield was evaluated in steady state, where the surface is roughened self-consistently due to the particle injections.  
However, it seems these differences did not affect the self-sputtering yield much and the results obtained in both studies are in good agreement in the energy range shown here.

The solid curve is a regression curve obtained from the standard Gaussian Process Regression\cite{RasmussenWilliams:2006qz} (GPR) based on the 37 data points (filled circles) shown in \refFig{F::EnergDepend}.
The gray band represents the standard deviation (SD) around the regression curve, representing the uncertainty.
The values of the GPR regression curve at selected normalized energies and the MD simulation data are presented in the supplementary material.
The threshold normalized energy, at which $Y = 1$, is $\mathcal{E} \simeq $ 132, based on the regression.
 
It is seen the sputtering yield $Y$ is an increasing function of the normalized incident energy $\mathcal{E}$ increases. However, the way it increases is not monotonic and the gradient of the curve changes around $\mathcal{E} \simeq 200$ and $10^3$.
The cause of this structure in the yield curve has not been clarified yet.

\begin{figure}[htbp]
\begin{center}
\includegraphics[width=0.8\columnwidth]{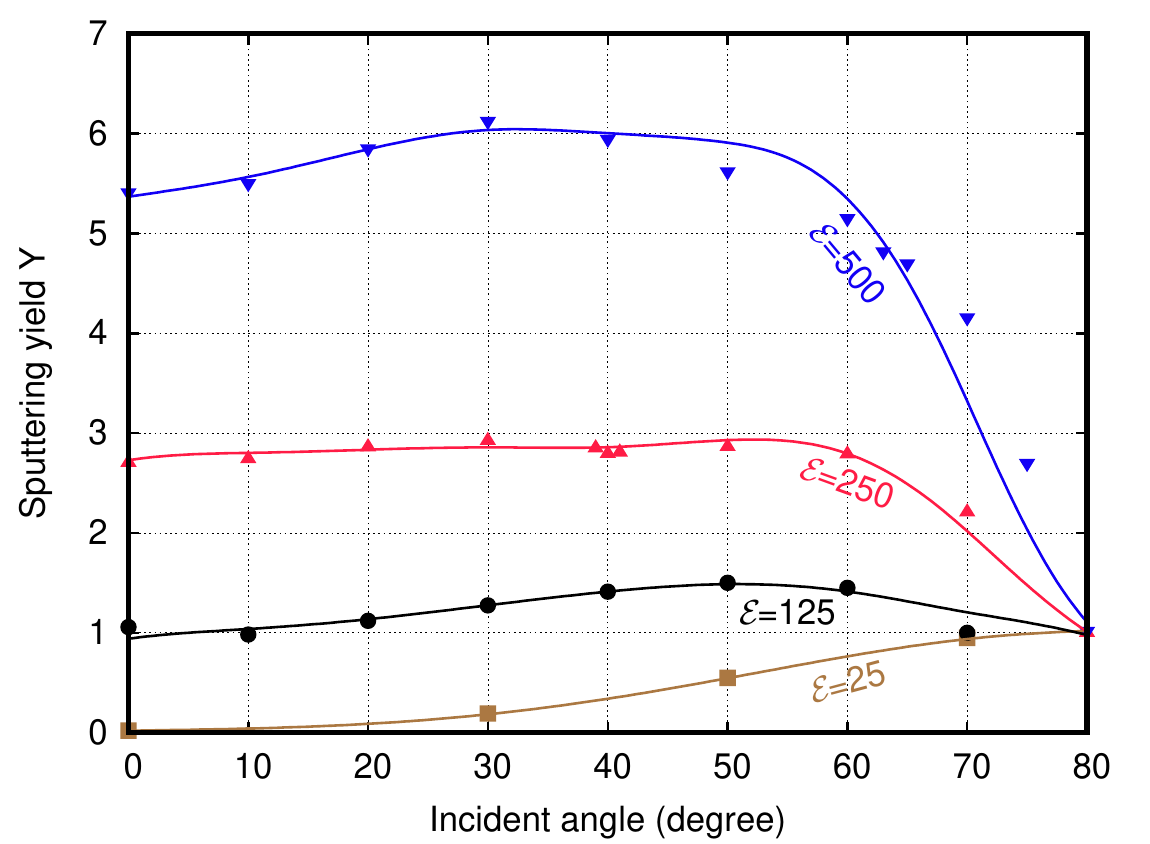}
\caption{Angular dependency of the self-sputtering yield of the LJ solid for four values of $\mathcal{E} =$ \numlist{25;125;250;500}.The solid curves are guides to the eye.
}
\label{F::AngDepend}
\end{center}
\end{figure}

The dependency of the self-sputtering yield on the angle of incidence is shown in Fig.~\ref{F::AngDepend} for different normalized incident energies. 
As seen here, the sputtering yield increases with the increasing angle of incidence $\alpha$ up to its maximum value and then decreases as the angle $\alpha$ further increases. The yield reaches $Y = 1$ at around $\alpha = 80^\circ$. The solid curves in \refFig{F::AngDepend} are guides to the eye.

\subsection{Comparison with experiments} 

Self-sputtering yields of various single-element materials for normal incidence have been obtained experimentally.
We now discuss how the experimentally obtained self-sputtering yield values of actual materials are compared with the self-sputtering yields of the LJ system given in \refFig{F::EnergDepend}.
To make such a comparison, we must identify the values of $\sigma$ and $\epsilon$, with which the LJ interatomic function of \refEq{Eq::LJpotential} approximately represents the atomic interactions of actual materials. 
Especially, as we discussed earlier, the self-sputtering yield of the LJ system for normal incident particles depends only on the normalized ion incident energy $\mathcal{E}$.
Therefore, to translate the experimentally obtained sputtering yield $Y$ as a function of the incident ion energy $E$, we only need to know the corresponding value of $\epsilon$ to convert $E$ to $\mathcal{E} = E/4\epsilon$. 

A simple way to estimate the corresponding value of $\epsilon$ for an actual material is to use the cohesive energy $E_C$, i.e., the energy per atom (or per mole) needed to separate the atoms consisting of the material into independent neutral atoms.
The cohesive energy has been obtained for most single-element materials.\cite{kittel:2004vq}
On the other hand, our numerical simulation has shown that the cohesive energy per atom of the fcc solid consisting of LJ particles is given by

\begin{equation}
    \label{Eq::LJCoheEnerg}
    E_C \simeq -8.6 \epsilon.
\end{equation}

at zero temperature and zero pressure. The result is consistent with the earlier estimate in Ref.~\citenum{ashcroft:1976aa}.
By comparing this relation with the cohesive energy of a material obtained either experimentally or by a reliable first-principles calculation, one can estimate the corresponding value of $\epsilon$. Even for a material that does not form an fcc crystal, one can formally define its corresponding $\epsilon$ value using its cohesive energy $E_C$ and the relation (\ref{Eq::LJCoheEnerg}) as a measure of its cohesiveness.

Similarly one can use the nearest neighbor distance of a material, which has been known for most single-element materials under certain thermodynamical conditions, \cite{kittel:2004vq} to estimate the corresponding value of $\sigma$ if the atomic interactions of the material can be approximately represented by the LJ potential. For the fcc solid consisting of LJ particles, our simulation shows that the nearest neighbor distance $d$, the lattice constant $l_a$, and $\sigma$ are related as

\begin{equation*}
d \simeq 0.71 l_a \simeq 1.1 \sigma
\end{equation*}

at zero temperature and zero pressure. This result is consistent with the earlier estimate in Ref.~\citenum{ashcroft:1976aa}.
The \refTab{T::LJparaEval} gives the estimated values of  $\epsilon$ and $\sigma$ for iron (\ce{Fe}), nickel (\ce{Ni}), copper (\ce{Cu}) and gold (\ce{Au}), based on their cohesive energies and nearest neighbor distances listed in Ref.~\citenum{kittel:2004vq}. Note that \ce{Ni}, \ce{Cu}, and \ce{Au} form fcc crystals whereas \ce{Fe} forms a body-centered cubic (bcc) crystal at standard temperature and pressure.  

\begin{figure}[htbp]
\begin{center}
\includegraphics[width=\linewidth]{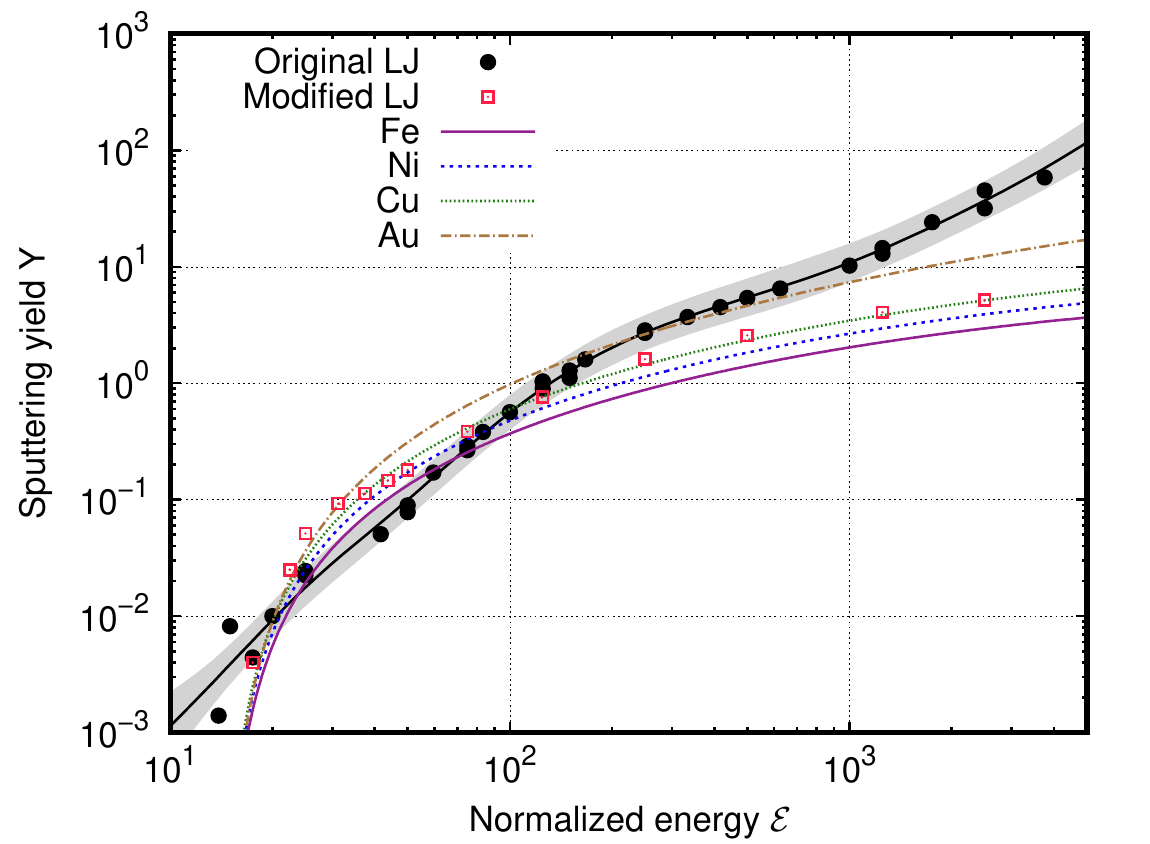}
\caption{Comparison between the self-sputtering yield curve of the LJ solid and those of \ce{Fe}, \ce{Ni}, \ce{Cu}, and \ce{Au} presented by the formula by Yamamura and Tawara.\cite{Yamamura_Tawara_ADNDT96} The red squares are the sputtering yields of the same solid based on the modified LJ potential $\tilde{U}$ of \refEq{E_PotModif}.
}
\label{F::EpsilonScale}
\end{center}
\end{figure}

Figure \ref{F::EpsilonScale} shows the comparison of the self-sputtering yield of the LJ solid (the same as Fig.~\ref{F::EnergDepend}) with those of \ce{Fe}, \ce{Ni}, \ce{Cu}, and \ce{Au} as functions of the normalized energy. The yield curves for \ce{Fe}, \ce{Ni}, \ce{Cu}, and \ce{Au} are fitting curves to experimental data, given by Yamamura and Tawara,\cite{Yamamura_Tawara_ADNDT96} where the incident ion kinetic energy $E$ was converted to the normalized energy $\mathcal{E}$ with \refEq{E_nomarlized_energy} and $\epsilon$ given in \refTab{T::LJparaEval}. 
The red squares are the self-sputtering yields obtained for an interatomic potential function that will be discussed in the following subsection.  

As seen in \refFig{F::EpsilonScale}, the self-sputtering yields for \ce{Fe}, \ce{Ni}, \ce{Cu}, and \ce{Au} are relatively close to those of the LJ solid when the normalized energy $\mathcal{E} \lesssim 100 $, where $Y <1$ and the deposition takes place. Considering the fact that the vertical axis is a log scale (and therefore magnifies a small difference for small numbers) and the LJ potential is typically considered as a simplistic approximation for the interatomic potentials of \ce{Fe}, \ce{Ni}, \ce{Cu}, and \ce{Au} atoms, the agreement in this normalized energy range is rather impressive. 
However, at higher energies, e.g., $\mathcal{E} \gtrsim 300$, the LJ solid exhibits systematically higher yield values. As discussed earlier, the repulsive part of the LJ potential at short distances is unrealistic, which causes this deviation from the self-sputtering of real materials.\cite{Mauchamp_Ni_EAM_21}
In the following subsection, we discuss a possible modification of the LJ potential at short distances. 

\begin{table}[htbp]
\caption{\label{T::LJparaEval}Evaluated values of $\epsilon$ and $\sigma$ for \ce{Fe}, \ce{Ni}, \ce{Cu} and \ce{Au} based on their cohesive energies and nearest neighbor distances listed in Ref.~\citenum{kittel:2004vq}.
}
\begin{ruledtabular}
\begin{tabular}{@{} l c c c c @{}}
 & \ce{Fe} & \ce{Ni} & \ce{Cu} & \ce{Au}\\
\midrule
$\epsilon$ (\si{\electronvolt}) & \num{0.50} & \num{0.52} & \num{0.41} & \num{0.45}\\
$\sigma (\si{\angstrom})$  & \num{2.27} & \num{2.28} & \num{2.37} & \num{2.64}
\end{tabular}
\end{ruledtabular}
\end{table}

\subsection{Potential modification} 
To lower the self-sputtering yield, we can reduce the collision cross section, allowing the incident particles to penetrate deeper and pass their momenta and energies to a larger number of the substrate atoms, some of which are located in a much deeper region. In this way, fewer recoiled atoms in the material may reach the surface and gain sufficient energy to leave the surface. To do so, we need to reduce the value of the potential function at short distances whereas the attractive part of the potential is unaffected by this modification, so that we can avoid modifying the mechanical and thermodynamic properties of the system as well as the self-sputtering yield at low energies.

The Ziegler-Biersack-Littmark (ZBL) potential \cite{Ziegler1985} is widely used to represent a short-range repulsive interaction among atoms. This function is defined for each pair of atomic species and has no adjustable parameter.
However, in this study, we use a different function that more or less shifts the repulsive part of the LJ potential toward a smaller particle separation, rather than a generic ZBL potential, which has a totally different functional form from the repulsive part of the LJ potential.  

\begin{figure}[htbp]
\begin{center}
\includegraphics[width=\linewidth]{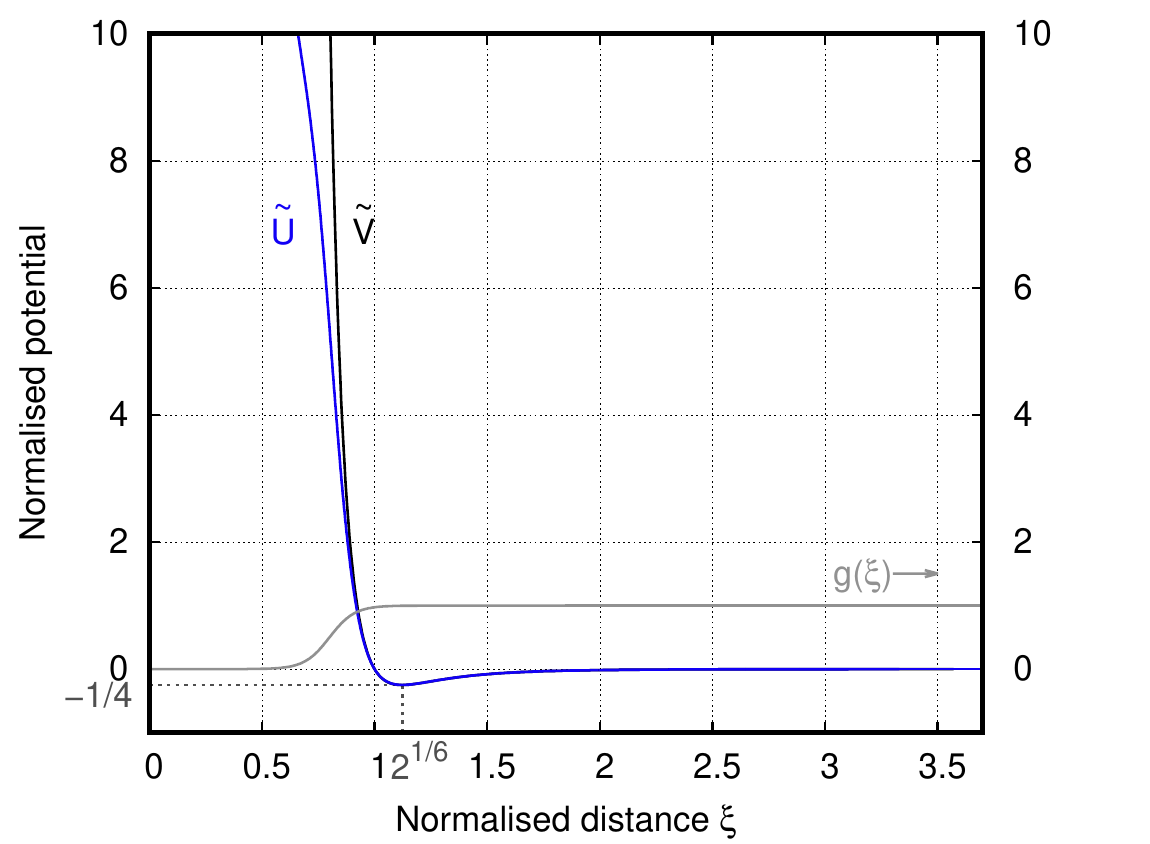}
\caption{Comparison between the normalized LJ potential function $\tilde{V}(\xi)$ and the normalized modified LJ potential function $\tilde{U}(\xi)$ at low energies.
The function $g(\xi)$ used to create $\tilde{U}$ is also plotted.
}
\label{F::LJpot_Norm}
\end{center}
\end{figure}

\begin{figure}[htbp]
\begin{center}
\includegraphics[width=\linewidth]{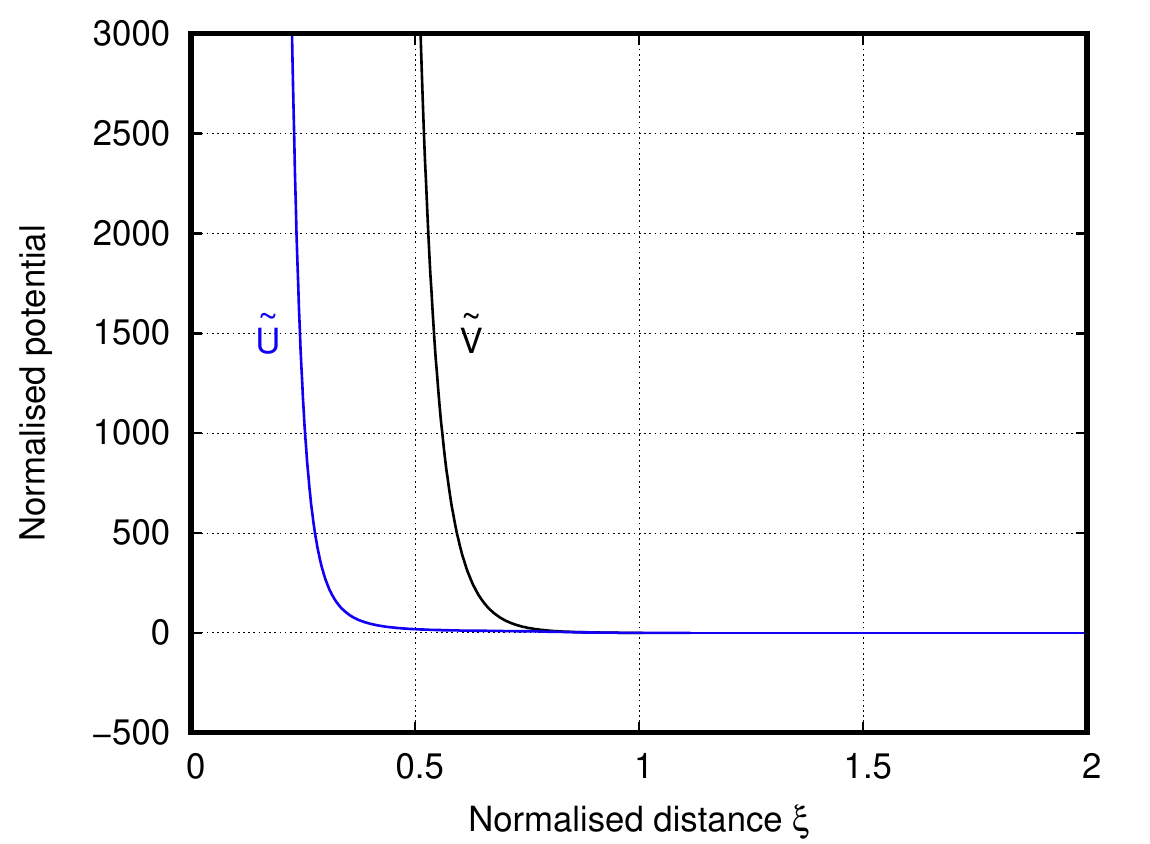}
\caption{Comparison between the normalized LJ potential function $\tilde{V}(\xi)$ and the normalized modified LJ potential function $\tilde{U}(\xi)$ at high energies.
}
\label{F::LJpot_Norm2}
\end{center}
\end{figure}

We define a modified normalized LJ potential function $\tilde{U}$ as

\begin{equation}
\tilde{U} (\xi) = g(\xi) \tilde{V}(\xi).   
\label{E_PotModif}
\end{equation}

with 

\begin{equation*}
g(\xi) =  A \tanh \left[ b \left( \xi - \xi_C \right) \right] +1 - A,
\end{equation*}

where $A$, $B$ and $\xi_C$ are free parameters.  
We selected $A = 0.4999$, $\xi_C = 0.8$ and $b = 9.0$ for this study such that the modified potential function would produce the self-sputtering yield curve close to that of a real material, as we shall discuss momentarily.
The function $g(\xi)$ serves as an activation function, i.e., $g(\xi) \simeq 1$ for $\xi \geq 1$, $g(\xi_C) \simeq  0.5$, and $0 < g(\xi) \ll 1$ for $0 < \xi \ll 1$, so that $\tilde{U}(\xi) \ll \tilde{V}(\xi)$ at short particle separation whereas  $\tilde{U}(\xi) \simeq \tilde{V}(\xi)$ at typical interatomic distances.
Figure~\ref{F::LJpot_Norm} shows the functional form of $g(\xi)$ (on the right vertical axis) as well as $\tilde{U}(\xi)$ and  $\tilde{V}(\xi)$ (on the left vertical axis).
The functions  $\tilde{U}(\xi)$ and $\tilde{V}(\xi)$ are hardly distinguishable below the potential value of 2 or so. Figure~\ref{F::LJpot_Norm2} shows the functional forms of $\tilde{U}(\xi)$ and $\tilde{V}(\xi)$ on a much larger scale. It is seen that the function $\tilde{U}(\xi)$ represents the particle interaction with much smaller particle "radii" when two particles collide at high kinetic energy.

The self-sputtering yields obtained from MD simulations based on this modified potential $\tilde{U}(\xi)$ are shown in \refFig{F::EpsilonScale} with red empty squares.
It is seen that they agree better with those of metals listed in this figure in a wide energy range.
The potential modification above was performed only to show the effect of reduced interaction radius (and therefore the reduced collision cross section) at high energy impact and we by no means attempt to claim that the functional form of \refEq{E_PotModif} is anything physical. As expected, the simulation-based self-sputtering yields were significantly reduced and brought to values close to those of real materials. On the other hand, at a lower energy range of $20 \lesssim \mathcal{E} \lesssim 50$, the self-sputtering yields of the modified LJ system are systematically higher than those of the original LJ system.  This is an opposite trend although the radius of impact in this energy range (i.e., the value of $\xi$ that satisfies $\mathcal{E} = \tilde{U}(\xi)$ for $\mathcal{E}$ in this energy range) is still smaller than that of the original LJ potential $\tilde{V}$. This may be related to the fact that, at shallow collisions, smaller radii of impact allow recoiled energetic particles to escape from the surface region to vacuum more easily by keeping wider open space among atoms located at the lattice sites.

\begin{figure}[htbp]
\begin{center}
\includegraphics[width=\linewidth]{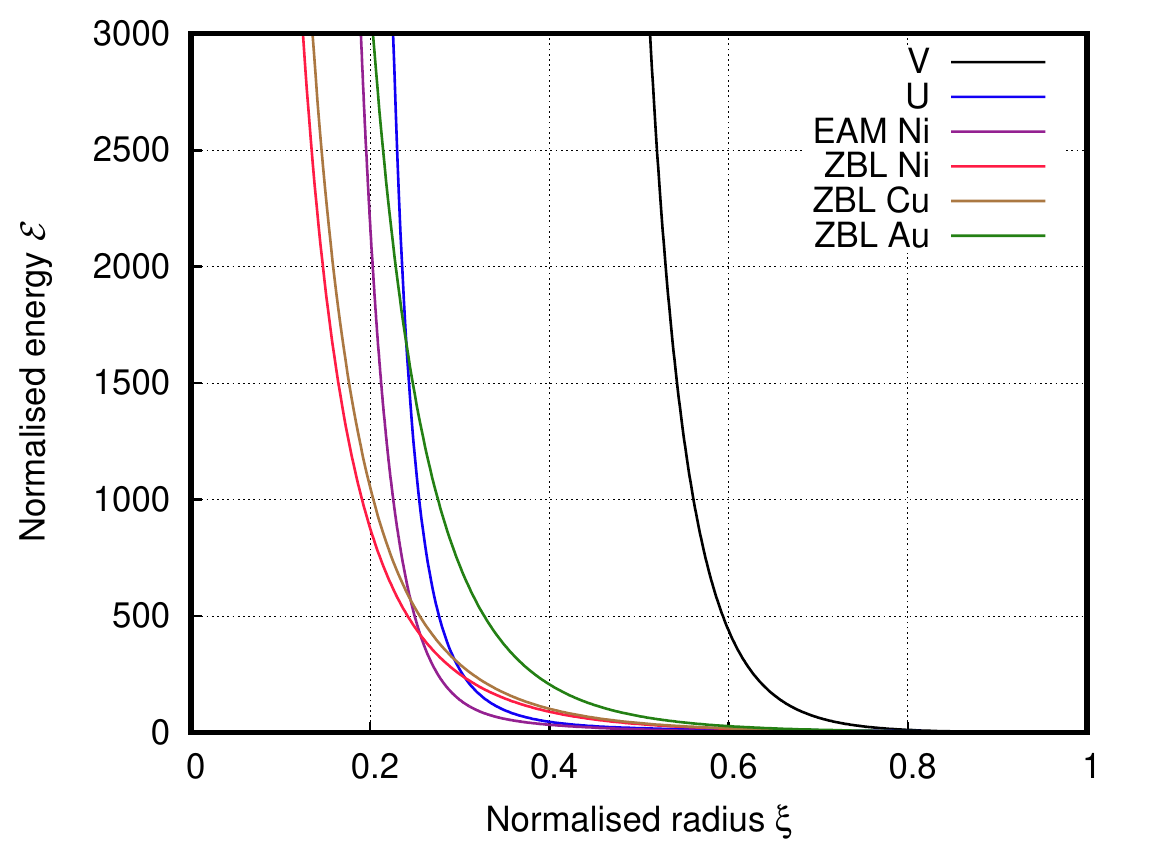}
\caption{Comparison among different potential functions at short separation. The black and blue curves represent the normalized LJ potential function $\tilde{V}(\xi)$ (denoted as V in the legend) and the normalized modified LJ potential function $\tilde{U}(\xi)$ (denoted as U in the legend). The purple curve is the normalized modified EAM potential function for \ce{Ni} used in Ref.~\citenum{Mauchamp_Ni_EAM_21}. 
The red, gold, and green curves are the normalized ZBL functions for \ce{Ni}, \ce{Cu}, and \ce{Au}, respectively. The ZBL function for \ce{Fe} overlaps with the ZBL function for \ce{Ni} in this scale. The normalization was performed with the values given in \refTab{T::LJparaEval}.
}
\label{F::LJpot_ZBL}
\end{center}
\end{figure}

We now compare the model function $\tilde{U}(\xi)$ with other known model potential functions at short particle separation. In \refFig{F::LJpot_ZBL}, the normalized LJ potential function $\tilde{V}(\xi)$ (denoted as $V$ in the legend) and the normalized modified LJ potential function $\tilde{U}(\xi)$ (denoted as $U$ in the legend) are compared with the normalized ZBL functions for \ce{Ni}, \ce{Cu}, and \ce{Au}, as well as the normalized modified EAM potential function for \ce{Ni} of Ref.~\citenum{Mauchamp_Ni_EAM_21}.
In Ref.~\citenum{Mauchamp_Ni_EAM_21}, this interatomic potential function for Ni is denoted as P3 and the MD simulation based on this potential function is shown to provide the self-sputtering yields of Ni in reasonable agreement with the experimental values in the incident ion energy range from \SIrange{100}{3000}{\electronvolt}. 
The ZBL function for \ce{Fe} overlaps with the ZBL function for \ce{Ni} in this scale and, therefore, it is not listed here. 
The normalization was performed with the values given in \refTab{T::LJparaEval}.
It is seen that 
the function $\tilde{U}(\xi)$ is closer to the EAM and ZBL functions listed here than the normalized LJ potential function $\tilde{V}(\xi)$. 
The sputtering yield is likely to depend on the balance between the repulsive interactions of atoms at close separation and the attractive interactions at intermediate separation.
Because the LJ potential is too simplistic to represent the attractive interactions for real atoms, we do not attempt to construct a repulsive interaction model for a better agreement between the simulation and experimental results within the framework of the LJ system of this study.

\section{Conclusions} 
\label{S::Concl}
MD simulations were performed to obtain the universal curve of the self-sputtering yield for the (100) surface of the LJ fcc crystal at low surface temperature as a function of the normalized incident particle kinetic energy for normal incidence.
We have also obtained the self-sputtering yields of the fcc LJ crystal as functions of the angle of incidence at some selected normalized incident energies.
Because the most stable crystalline structure of such a system is the fcc crystal and its self-sputtering yield depends only on the normalized incident energy and the angle of incidence, the sputtering yield curves presented here serve as the fundamental reference data for the LJ potential.
We also compared the self-sputtering yield of the LJ solid with those of \ce{Fe}, \ce{Ni}, \ce{Cu}, and \ce{Au} as functions of the incident ion energy for normal incidence.
It should be noted that \ce{Ni}, \ce{Cu}, and \ce{Au} form fcc crystals whereas \ce{Fe} forms a bcc crystal at standard temperature and pressure.
Nevertheless, the agreement is reasonable at low ion incident energy where deposition typically takes place.
At higher incident energy, however, the self-sputtering yield of the LJ solid differs significantly from those of the actual materials.
This is because, as pointed out by various previous studies, the repulsive interactions of the LJ potential do not represent those of actual atoms at short distances.

If the incident particle is a different species from those of the surface material, the sputtering yield depends on more parameters characterizing the difference in species, in addition to the ion incident energy, even if all particles interact with the LJ potentials. 
The extension of the current study to such a system may serve as a reference for more general sputtering phenomena of real materials and will be published separately. 

\section*{Supplementary Material}
Three supplementary text files are provided with this paper.
The file ``SuppMat\_EnergDep.txt'' lists the self-sputtering yields obtained using the original 12-6 LJ potential function and the modified LJ potential function as a function of the normalized energy $\mathcal E$ (Figs.~\ref{F::EnergDepend} and~\ref{F::EpsilonScale}).
The file ``SuppMat\_EnergDep\_GPR.txt'' lists the predicted self-sputtering yield obtained using Gaussian Process Regression (GPR), as well as the upper and lower uncertainty (Figs.~\ref{F::EnergDepend} and~\ref{F::EpsilonScale}).
The file ``SuppMat\_AngDep.txt'' contains the self-sputtering yields as a function of the incident angle (\refFig{F::AngDepend}).

\section*{Acknowledgements}
The authors are grateful to Prof. \v{S}tefan Matej\v{c}\'{i}k of Comenius University for helpful discussion. 
This work was partially supported by Japan Society of Promotion of Science (JSPS) Grant-in-Aid for Scientific Research (S) 15H05736 and (A) 21H04453, JSPS Core-to-Core Program JPJSCCA2019002, and Osaka University International Joint Research Promotion Programs (Type A).

\section*{Data availability}
The data that support the findings of this study are available from the corresponding author upon reasonable request.

%

\end{document}